\documentclass[10pt, downdrafts, twocolumn]{IEEEtran}  

\usepackage{subfigure}
\usepackage{pdfpages}
\usepackage{diagbox}
\usepackage{colortbl}
\definecolor{mygray}{gray}{0.85}
\usepackage{array}
\usepackage{caption}
\usepackage[linesnumbered,boxed,ruled,commentsnumbered]{algorithm2e}

 \usepackage{eqparbox}

\usepackage{amsmath,bm}
\usepackage{cite}
\usepackage{amsmath,amssymb,amsfonts}
\usepackage{amsthm}
\usepackage{graphicx}
\usepackage{textcomp}
\usepackage{xcolor}

\usepackage{hyperref}
\hypersetup{%
  bookmarks=true
}

\begin{document}
\title{
Edge Intelligence for Autonomous Driving in 6G Wireless System: Design Challenges and Solutions
}

\author{
\IEEEauthorblockN{Bo~Yang, \IEEEmembership{Member,~IEEE}, Xuelin Cao, Kai Xiong, Chau Yuen, \IEEEmembership{Senior Member,~IEEE},  Yong Liang Guan, \IEEEmembership{Senior Member,~IEEE}, Supeng Leng, \IEEEmembership{Member,~IEEE},  Lijun Qian, \IEEEmembership{Senior Member,~IEEE}, and Zhu Han,~\IEEEmembership{Fellow,~IEEE}
}
\vspace{-0.25cm}
}

\maketitle

\begin{abstract}
In a level-5 autonomous driving system, the autonomous driving vehicles (AVs) are expected to sense the surroundings via analyzing a large amount of data captured by a variety of onboard sensors in near-real-time. As a result, enormous computing costs will be introduced to the AVs for processing the tasks with the deployed machine learning (ML) model, while the inference accuracy may not be guaranteed. In this context, the advent of edge intelligence (EI) and sixth-generation (6G) wireless networking are expected to pave the way to more reliable and safer autonomous driving by providing multi-access edge computing (MEC) together with ML to AVs in close proximity. To realize this goal, we propose a two-tier EI-empowered autonomous driving framework. In the autonomous-vehicles tier, the autonomous vehicles are deployed with the shallow layers by splitting the trained deep neural network model. In the edge-intelligence tier, an edge server is implemented with the remaining layers (also deep layers) and an appropriately trained multi-task learning (MTL) model. In particular,  obtaining the optimal offloading strategy (including the binary offloading decision and the computational resources allocation) can be formulated as a mixed-integer nonlinear programming (MINLP) problem, which is solved via MTL in near-real-time with high accuracy. On another note, an edge-vehicle joint inference is proposed through neural network segmentation to achieve efficient online inference with data privacy-preserving and less communication delay. Experiments demonstrate the effectiveness of the proposed framework, and open research topics are finally listed.
\end{abstract}

\section{Introduction}
Different from the conventional cellular networks, the next-generation wireless networks, with various labels such as beyond fifth-generation (B5G) or sixth-generation (6G), are expected to provide a connection for vehicular networks with low latency but high reliability ubiquitously via artificial intelligence (AI)~\cite{6G}. To prevent accidents caused by human driving errors and to carry out emissions reduction, automated driving systems have been developed by tightly integrating with promising technologies, such as surround sensing, object detection and tracking, control decision-making, and wireless offloading via multi-access edge computing (MEC)~\cite{{self-driving-survey},{ZY_TVT2}}. During the past decades, machine learning (ML), especially deep learning (DL), has become a promising technology due to its superiority in extracting inherent features from large-scale and high-dimensional datasets, which are collected from embedded sensors such as camera sensors, LiDAR sensors, radar sensors~\cite{ML_road}. By feeding the inferred results to the electronic control unit (ECU) of a vehicle, the AVs can dynamically adjust the vehicle's velocity, brake, and steering according to the surrounding conditions to achieve autonomous or cooperative adaptive cruise control (CACC). Nevertheless, the inferring accuracy usually depends on the input data and capability of the deployed DL model. For example, the object is difficult to be identified via the convolutional neural network (CNN) model if the input images data is blurry or if the CNN model has not been sufficiently trained~\cite{YB_WCL}. 

In recent years, Internet-of-Vehicle (IoV) networks undergo a paradigm shift from connected vehicles to connected intelligence. The future of vehicular communications may be a mashup of dedicated short-range communications (DSRC) and cellular vehicular-to-everything (C-V2X, also LTE Rel.16+). Today the fight among vehicular communication standards is between ``802.11p/DSRC" and ``C-V2X" ~\cite{XK_ITS}.  802.11p has the advantage that 5.9 GHz has been approved by the Department of Transportation and the European Commission, and the technology is ready for deployment for basic short-range communications. For ``Driver-less V2X", the situation is different. Due to the massive data anticipated in MEC, 802.11p will likely not be sustainable without significant amendments to the current standard. In this context, the innovative aspects anticipated via C-V2X within B5G/6G will play a key role. 

\begin{figure*}[t] 
            \centering
            \captionsetup{font={small }}
            \includegraphics[width=5.9in, height=3.0in]{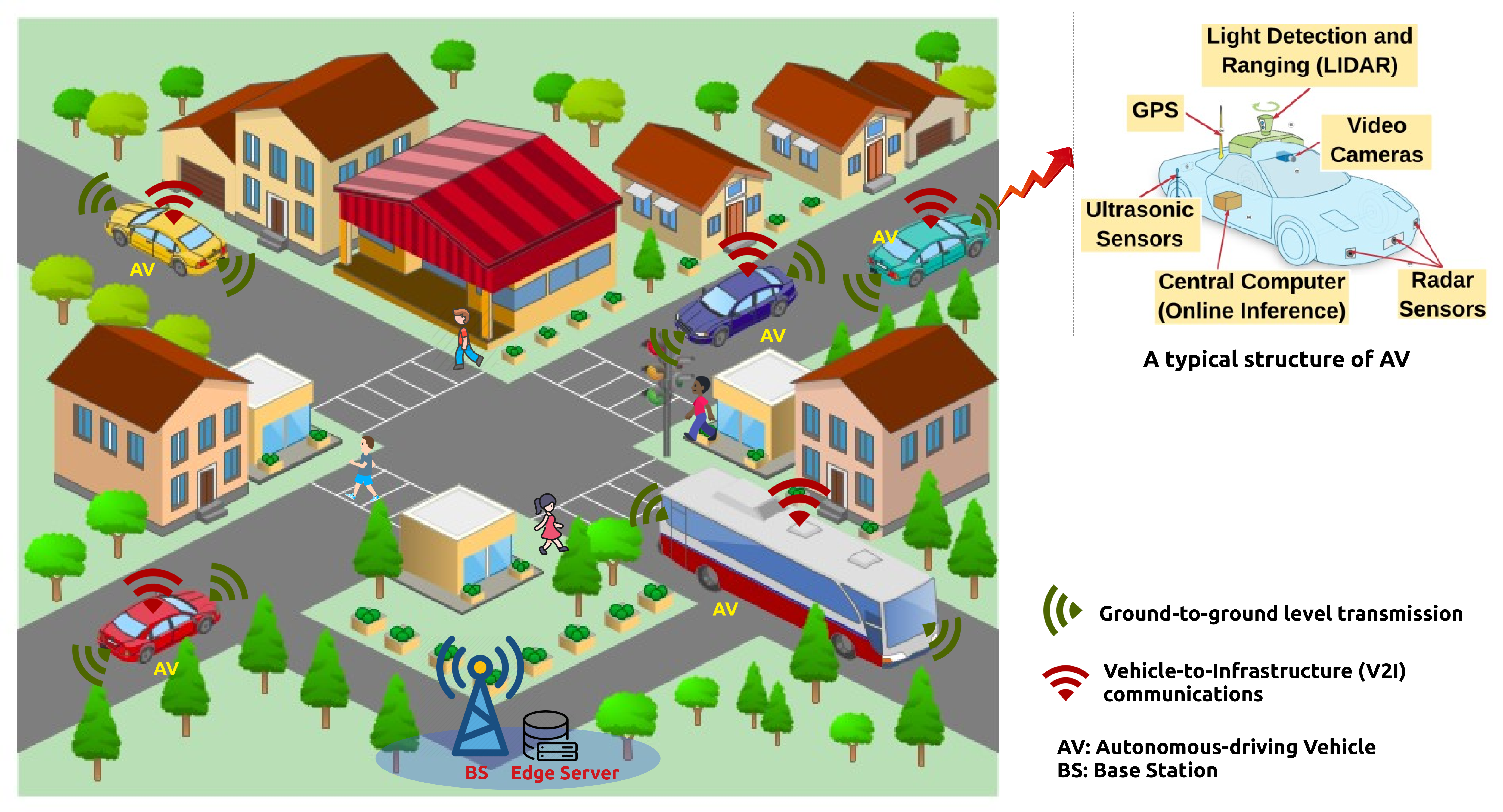} 
            \caption{An illustrative scenario of autonomous driving system.}
            \label{fig:1}\vspace{-0.2cm}
\end{figure*}

As a promising application, the thriving autonomous driving has led to an upsurge of MEC and AI, which drives the prosperity of \textit{edge intelligence} (EI) for greatly facilitating daily life~\cite{{ZY_IOT},{ZY_Network},{ZY_WCM}}. In this context, EI enables autonomous driving vehicles (AVs) to sense and react to the surroundings accurately by offloading the collected data to the powerful edge server co-located with the base station (BS), as highlighted in Fig.~\ref{fig:1}. By deploying EI-based computing service at the edge server, inference computing of AVs can be achieved with higher accuracy and lower latency.
However, many inherent issues associated with communication and computing under limited bandwidth and computation resources, data privacy and security, present stumbling blocks toward the envisioned goal of autonomous driving in 6G era. Motivated by the above-mentioned appealing characteristics, a flurry of research activities combining 6G with EI in the autonomous driving system designs has been sparked.

In this vein, this article aims to explore the EI challenges in autonomous driving, investigate recent contributions and advances, and then propose solutions. The major contributions and organization of this article are summarized as follows. We overview the efforts that have been made on edge intelligence in vehicular networks and pose the design challenges in Section~\ref{sec2}. In Section~\ref{sec3}, we present the developed two-tier EI-empowered autonomous driving framework, which outperforms the existing works in the following two folds: 1) {achieving the intelligent offloading strategy via multi-task learning in near-real-time}, and 2) {achieving the joint inference with privacy-preserving and less communication delay through neural network segmentation}. In particular, a multi-task learning model is deployed at the edge server to infer the optimal offloading decision of the vehicles and the computation resource allocation of the edge server with high accuracy in near-real-time, augmented by a
neuron network segmentation method to achieve efficient online inference dynamically with privacy-preserving and less latency raised. A case study is presented to demonstrate the proposed framework and then evaluated via experiments in Section~\ref{sec4}, followed by listing some critical open research topics and future guidelines in Section~\ref{conclusion}.

\section{Where Edge Intelligence Meets Autonomous Driving: Applications and Challenges}
\label{sec2}
In this section, we list some EI applications in autonomous driving and discuss the technical challenges raised.

\subsection{Applications of EI in Autonomous Driving}
There has been lately increasing research interest in incorporating AI technology (e.g., ML and DL) into edge computing, known as EI, which can help achieve efficient computing and intelligent decision making to keep the roads safe. In the following, we list some fundamental computation-intensive applications aided by EI in autonomous driving, such as object detection, traffic flow prediction, and path planning.
 
\subsubsection{Object Detection}  
Real-time video analytics has been widely applied in autonomous driving (see Fig.~\ref{fig:1}), where the data captured by the cameras and other onboard sensors is analyzed via the AI techniques. To achieve safe and efficient driving (e.g., reduce accidents and decrease traffic congestion), autonomous vehicles should identify the objects timely and accurately according to the surroundings. However, executing this kind of delay-sensitive tasks via DL generally requires high computation resources onboard and incurs high bandwidth consumption by uploading the tasks to the remote cloud centers. To address the issues and make autonomous driving possible, EI enables the self-driving vehicles to move some of the video analysis to the edge servers, which are usually near the data sources and can improve the inference accuracy.

\subsubsection{Traffic Flow Prediction}  
With the ever-increasing human demands on the intelligent transportation system and the adoption of autonomous vehicles, timely and accurate acquisition of traffic flow information will lay out the foundation for many location-related applications, e.g., navigation. However, massive volumes of environmental data with various sources will seriously saturate the onboard storage and become too complex to satisfy the requirements with the traditional computing techniques and infrastructures~\cite{feng}. To meet this challenge, EI represents a trend of ``driverless revolution on big data". Specifically, by combining AI with advanced edge computing techniques, the computation resources can be expanded to the edge server to achieve state-of-the-art performance for hierarchical features learning from the high-dimensional dataset and satisfy real-time requirements of the time-consuming tasks. 
 
\subsubsection{Intelligent Decision}
Recent years have shown remarkable attention on intelligent decision in autonomous driving,  e.g., online path planning, which aims to improve road safety by avoiding collision with the surrounding obstacles, vehicles, and infrastructure. With feasible path planning, AVs can make reasonable decisions on the critical maneuvers (such as braking, turning, and overtaking.) according to the surroundings by feeding the inferring results into the onboard ECU or the CACC system. In practice, intelligent decision can be formulated as an optimized decision-making problem taking into account the vehicle dynamics, which, however, makes the decision hardly be achieved in time~\cite{feng}. Notably, this becomes a concern increasingly when the B5G/6G paradigm is taken into account. To meet this concern, EI can be a revolutionary breakthrough by inferring an optimal decision directly in near-real-time via a well trained DL model deployed at the edge~\cite{YB_TMC}.                                                                                                                                                                                                                                                                                                                                                                                                                                                                                                                                                                                                                                                                                                                                                                                                                                                                                                                                                                                                                                                 Different from the conventional cloud computing-based methods, e.g., Google Map, with the aid of AI functioning on the edge, EI can significantly decrease the wireless communication delay by allowing AVs to upload the collected data (such as vehicle conditions and surroundings data) to the edge server. As a result, the AVs may obtain the inferring results within a limited time, and thus an intelligent, timely, and reliable decision is prone to be achieved at AVs, especially when the surrounding conditions change dramatically.
                                                                                                                                                                                                                                                                                                                                                                                                                                                                                                                                                                                                                                                                                                                                                                                                                                                                                                                                                                                                                                                                                                                                                                                                                                                                                                                                                                                                                                                                                                                                                                                                                                                                                                                                                                                                                                                                                                                                                                                                                                                                                                                                                                                                                                                                                                                                                                                                                                                                                                                                                                                                                                                                                                                                                                                                                                                                                                                                                                                                                                                                                                                                                                                                                                                                                                                                                                                                                                                                                                                                                                                                                                                                                                                                                                                                                                                                                                                                                                                                                                                                                                                                                                                                                                                            
\subsection{Technical Challenges}

 \subsubsection{{Infeasible Sensing}}                                                                                                                                                                                                                                                                                                                                                                                                                                                                                                                                                                                                                                                                                                                                                                                                                                                                                                                                                                                                                                                                                                                                                                                                                                                                                                                                                                                                                                                                                                                                                                                                                                                                                                                                                                                                                                                                                                                                                                                                                                                                                                                                                                                                                                                                                                                                                                                                                                                                                                                                                                                                                                                                                                                                                                                                                                                                                                                                                                                                                                                                                                                                                                                                                                                                                                                                                                                                                                                                                                                                                                                                                                                                                                                                                                                                                                                                                                                                                                                                                                                                                                                                                                                                                      In practice, due to the intrinsic features of the embedded sensors, they usually have restricted perception capacities, which lead to infeasible sensing of surroundings. Specifically, as a primary sensing method, the main drawback of visual-based object detection is that the inference performance (e.g., the accuracy of multi-class classification) usually relies on the captured images quality. The image quality can be affected by the surroundings, such as light conditions, weather conditions, image resolution, and the distance between camera and object. According to National Transportation Safety Board (NTSB) accident report, a Uber self-driving testing vehicle struck a pedestrian who was dressed in dark clothing and walked a bicycle across the road in Arizona at about 9:58 p.m. on March 18, 2018. The bicycle did not have any side reflectors, and the roadway lighting did not directly illuminate the roadway section. As a result, the pre-trained deep learning model mistakenly classified the pedestrian as an unknown object first, then as a vehicle, and finally, as a bicycle with varying expectations of the future travel paths. It is observed from this accident that the decision-making by a single kind of sensors (e.g., the cameras in Uber's testing case) alone has tended to introduce serious accidents. To avoid this, a promising candidate method is to fuse alternate sensing modalities for perception, e.g., fusing from the LiDAR sensors, the sensors implemented at the intelligent traffic lights, and querying the sensors from other vehicles. This kind of technology is known as sensor fusion, 

\subsubsection{Trade-off between Reliability and Latency}
 The inference performance generally includes not only the inference accuracy but also the inference delay. Since the inference accuracy of a pre-trained DL model is mainly determined by the input data quality and the model capability, the inference accuracy can be degraded due to the ``bad quality" data and shallow neural networks model. Besides, an additional communication delay is introduced by offloading the tasks from the mobile device to a more powerful edge server with a deeper neural network model. Nevertheless, the channel dynamics may sometimes disable the offloading. Even if not, the introduced delay could be fatal to delay-sensitive applications. Therefore, there exists a \textit{trade-off} between inference accuracy and latency, which makes careful consideration on the design of edge intelligence inevitable~\cite{YB_WCL}. In order to achieve a comfortable balance between the reliability and the latency raised, it becomes necessary to decrease the wireless transmission delay between the devices and the edge server. For example, the achievable rate of the wireless link can be improved via some promising technologies in future 6G networks, such as mmWave transmission, massive MIMO, and holographic reconfigurable intelligent surfaces. 
 On another note, the data can further be offloaded to other mobile devices via the device-to-device links with a shorter distance. In the considered vehicular networks, the vehicles may directly offload the data to their neighboring vehicles via cellar-V2V or even PC5 interfaces.
 
\subsubsection{{Limited Resources}}
Compared to a large amount of powerful graphics processing units (GPUs) and central processing unit (CPUs) integrated at the cloud, the edge servers usually can hardly bear a massive volume of offloading requests from the devices due to the constraint of computation, caching, and power resources on edge servers, and limited communication bandwidth. In this context, joint optimization of offloading decisions and allocation on the limited resources available at the edge server plays a particularly important role in edge intelligence. 
 
\subsubsection{{Data Security and Privacy}}
In many application domains of edge intelligence, data security and privacy issues are critical because the mobile devices' data required to be processed and inferred might carry much private-sensitive information that may not want to be captured~\cite{{ZY_TVT},{EI}}. Take autonomous driving as an example, where the AVs capture a vast amount of images containing privacy information. Suppose that this kind of information is directly offloaded to the edge server to be processed, the user privacy may leak. In general, this issue may occur in both the model training and inference. To address this issue, performing simple processing (or training) at the device first and then uploading the intermediate features to the edge server become promising.

\begin{figure*}[t] 
            \centering
           \captionsetup{font={small }}
            \includegraphics[width=5.0in, height=3.25in]{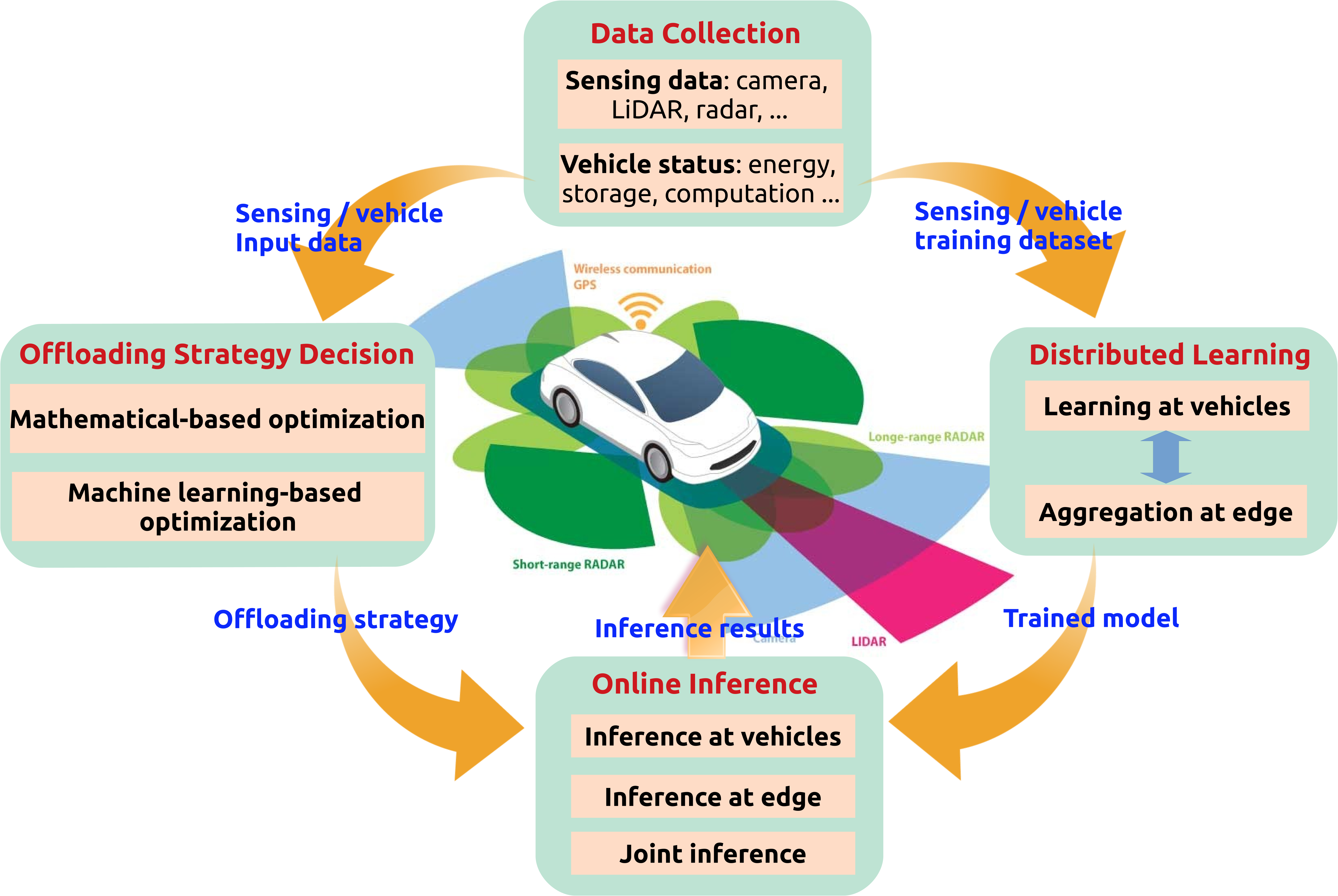} 
            \caption{Illustration of the four functionality modules.}
            \label{fig:2}\vspace{-0.2cm}
\end{figure*}

\section{Main Functionalities in Autonomous Driving Edge Intelligence}
\label{sec3}
In this section, we discuss how edge intelligence supports the key functionalities of autonomous driving and list four functionality modules,  as illustrated in Fig.~\ref{fig:2}.  

\subsection{Data Collection Module}
As a fundamental building block in the autonomous driving system, the data collection module collects the data via the onboard sensors. In general, the collected data can be roughly classified into two categories: sensing data and vehicle status data. In particular, the sensing data includes the data collected from the onboard sensors, such as video cameras, LiDAR sensors, and radar sensors. The vehicle status data usually reveals the vehicles' computation resources, e.g., the CPU/GPU processing capability, energy power, and storage capabilities.

\subsection{Distributed Learning Module}
Distributed learning (e.g., the federated learning) via the local datasets has gained considerable attention for providing an EI service for autonomous driving with privacy-preserving~\cite{EI}. In particular, distributed learning is operated at the edge devices based on their collected datasets. Then the trained local model parameters are uploaded to the edge server for aggregation. This procedure repeats many times until the model converges. In this context, AVs can collaboratively train a shared model using real-time generated mobile data. 
Although the distributed learning performed by massive edge devices is sometimes time-consuming due to the limited communication bandwidth, it is envisioned that distributed learning architectures will still get a considerable boost in autonomous driving since more advanced communication technologies are being used under the future B5G/6G networks~\cite{6G}.

\subsection{Offloading Strategy Decision Module}
Offloading strategy decision making within the critical time window is crucial for safety achievement in autonomous driving networks.
Traditionally, offloading strategy decision making includes two aspects: offloading decision and computation resources allocation. The offloading strategy decision making can be formulated as an optimization problem with performance metrics in latency and energy consumption. For the general self-driving scenarios, offloading strategy decision making involves solving a mixed integer programming problem that is usually NP-hard and difficult to solve via conventional mathematical techniques near-real-time~\cite{ZY_TVT2}. One promising approach is to apply deep learning for solving this kind of optimization problem by training a deep learning model to learn the mapping between the problem input parameters and the optimal solution. 
 
\subsection{Online Inference Module}
As the final step in the edge learning loop (as shown in Fig.~\ref{fig:2}), the driving maneuvers are generated by the online inference module based on the new input data.
On the one hand, for the local inference mode (i.e., the inference is made at the vehicles), the inference performance mainly depends on the vehicle's computing capability, which is power limited. On the other hand, introducing edge inference could provide low-latency services for AVs by feeding the data into the trained DL model deployed at the edge server. However, edge inference still faces many challenges, e.g., data privacy and limited computation capability, which can hardly fully bear and store the sizeable trained ML model.


\begin{figure*}[t] 
            \centering
            \captionsetup{font={small }}
            \includegraphics[width=6.4in, height=3.45in]{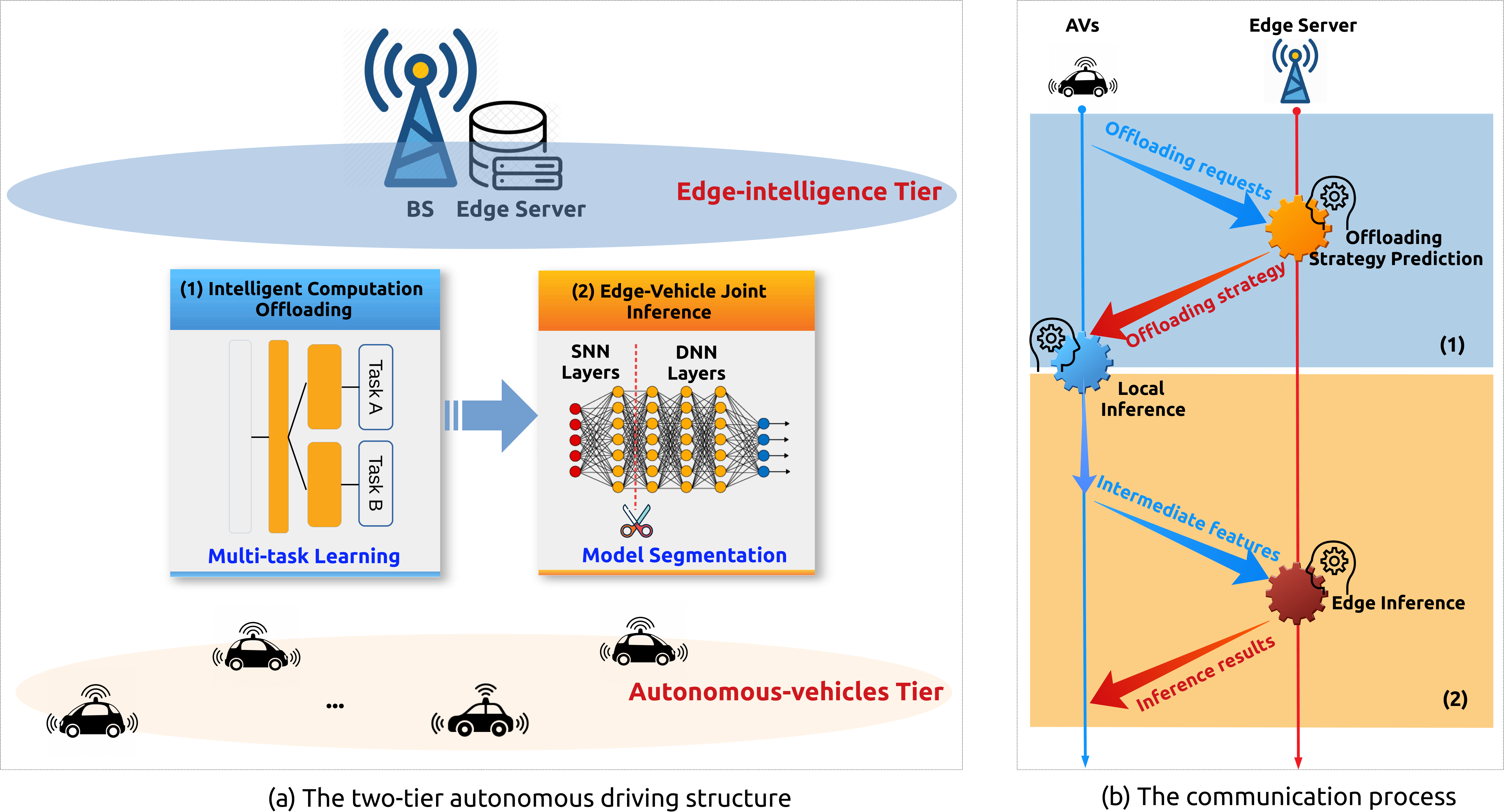} 
            \caption{An illustrative two-tier edge intelligence structure for autonomous driving system.}
            \label{fig:3}
\end{figure*}

\section{Two-Tier Edge Intelligence-Empowered Autonomous Driving Framework}
\label{sec4}
Motivated by the preceding considerations, we materialize EI-empowered object detection for autonomous driving and propose in this article a two-tier (including the AV-tier and the EI-tier) autonomous driving framework with a set of design guidelines, as shown in Fig.~\ref{fig:3}. In the following, we shall discuss specific components of the proposed framework and provide a concrete application example to illustrate the paradigm shift of autonomous driving.

\begin{figure*}[t] \vspace{-0.0cm} \hspace{-0.0cm}
            \centering
            \captionsetup{font={small }}
            \includegraphics[width=6.88in, height=5.05in]{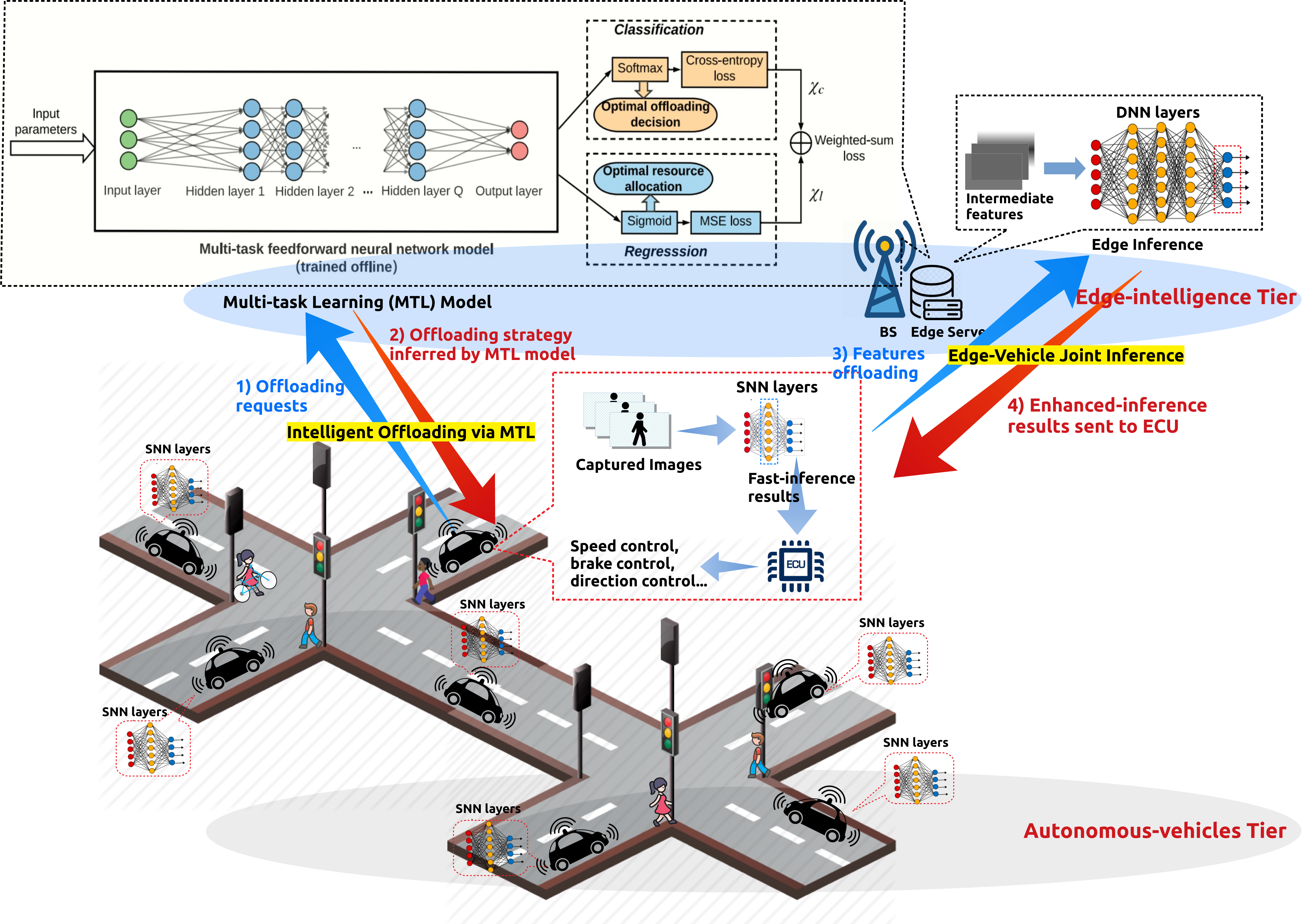} 
            \caption{A case study on the proposed two-tier autonomous driving framework.}
            \label{fig:4}\vspace{-0.2cm}
\end{figure*}\vspace{-0.0cm} \vspace{-0.0cm}

\subsection{Intelligence Offloading via Multi-task Learning}
In the EI-based autonomous driving system, we formulate the binary offloading decision-making (i.e., offload or not) and the computational resources allocation (denoting how many computational resources can be allocated to each vehicle) as a mixed-integer nonlinear programming (MINLP) problem, which is generally NP-Hard and challenging to solve. 
In particular, some factors (such as the number of vehicles, the computational capability of vehicles, and the channel conditions.) may vary over time, so the conventional relaxation-based optimization procedure must be executed repeatedly on solving the MINLP each time the parameters change. Therefore, high computational complexity is incurred due to numerical iterations, and the solutions are often sub-optimal and would not scale well~\cite{{YB_TMC}}. This is increasingly concerned when B5G/6G and IoT paradigms are taken into account~\cite{6G}.
 To meet the coming challenges, one promising approach is to apply deep learning for solving this kind of NP-hard optimization problem by training a deep learning model to `learn' the mapping between the problem input parameters and the optimal solution. In this paper, we build a multi-task learning (MTL) based framework to infer the solutions more efficiently with high accuracy, as illustrated in Fig.~\ref{fig:3}(a). In particular, a deep neural network model is designed and trained offline at the BS to obtain the mapping relationship from the input parameters to the output solutions. Therefore, on receiving the offloading requests from the AVs, the BS can directly infer the optimal offloading strategy decision in near-real-time by performing feedforward calculation via the MTL model without iterations, as shown in Fig.~\ref{fig:3}(b).

\subsection{Edge-Vehicle Joint Inference}
In the autonomous driving system, the online inference can be classified into three modes: local-inference mode, edge inference mode, and joint inference mode. For the local inference, the vehicles perform the inference themselves. The inference performance mainly depends on the available computational capability (computation resources and storage space) of the vehicles, which, however, is usually very limited~\cite{YB_WCL}. For the edge inference, the raw data must be uploaded to the edge server via the wireless uplink. We note that there exist two potential hazards associated with this approach. One is data privacy due to data fully uploading to the edge server. The other is that considerable communication delay is introduced once many vehicles upload their raw data with large size and quantity via the bandwidth-limited uplink channel. Motivated by the challenges in the previous two inference modes, the edge-vehicle joint inference becomes promising by dividing the trained deep neural network (DNN) model into two parts: the shallow neural network (SNN) layers at the vehicle side, and the DNN layers at the edge side. Therefore, the vehicles only need to upload the intermediate parameters to the edge server to reduce the end-to-end latency with joint-inference. In practice, the limited latency gains can be achieved due to the data amplification effect, which indicates that the size of intermediate output data of the deep neural networks is larger than that of the input data. In this context, choosing an appropriate splitting point while avoiding the data amplification effect becomes critical, and thus the size of uploaded parameters could be far less than the size of raw data by designing the neural networks model appropriately.

\begin{figure*}[t]
\centering
  \captionsetup{font={footnotesize }}
\subfigure[]{
\includegraphics[width=3.31in, height=2.15in]{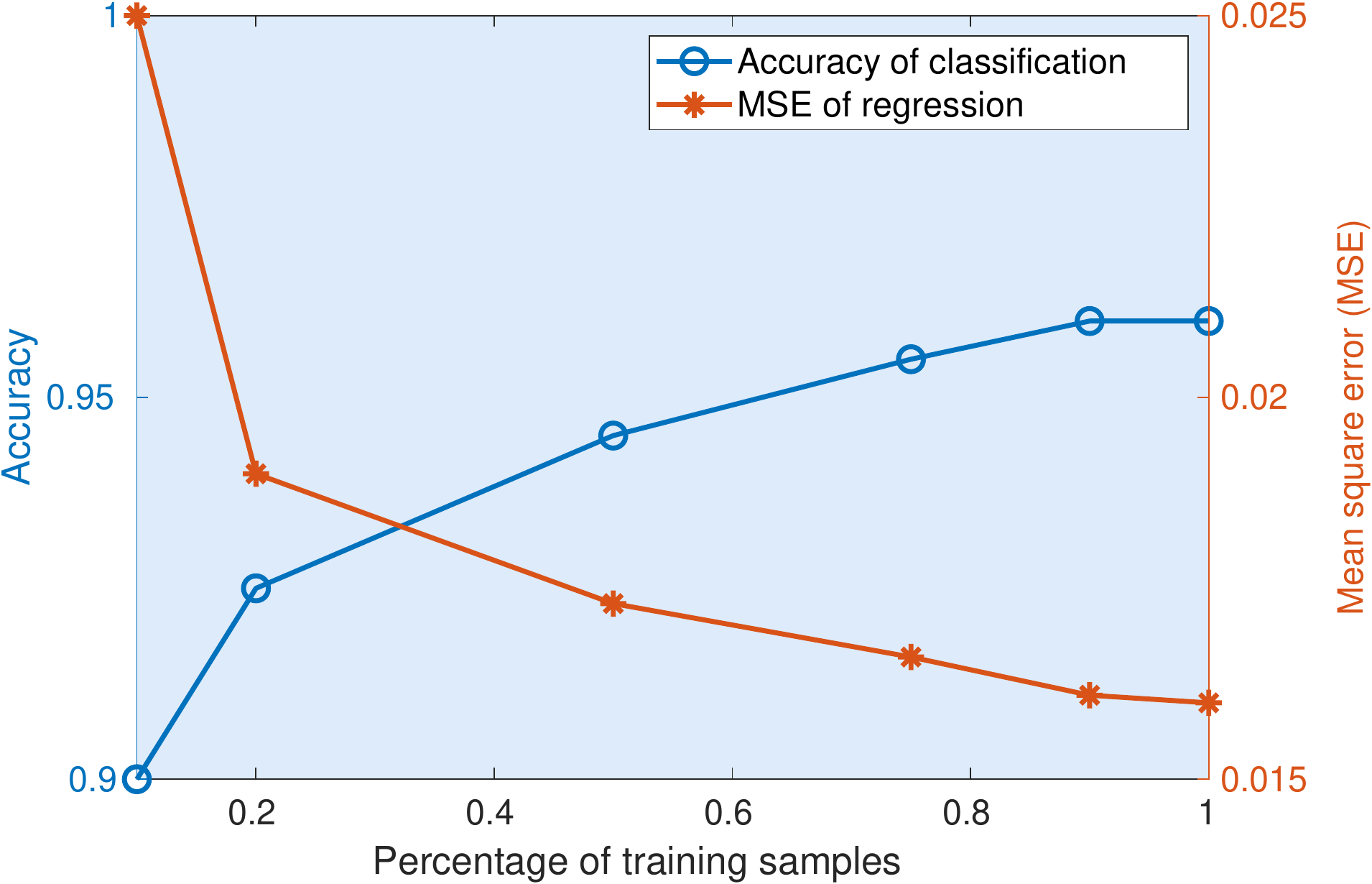}}
\hspace{0.2mm}
\subfigure[]{
\includegraphics[width=2.9in, height=2.15in]{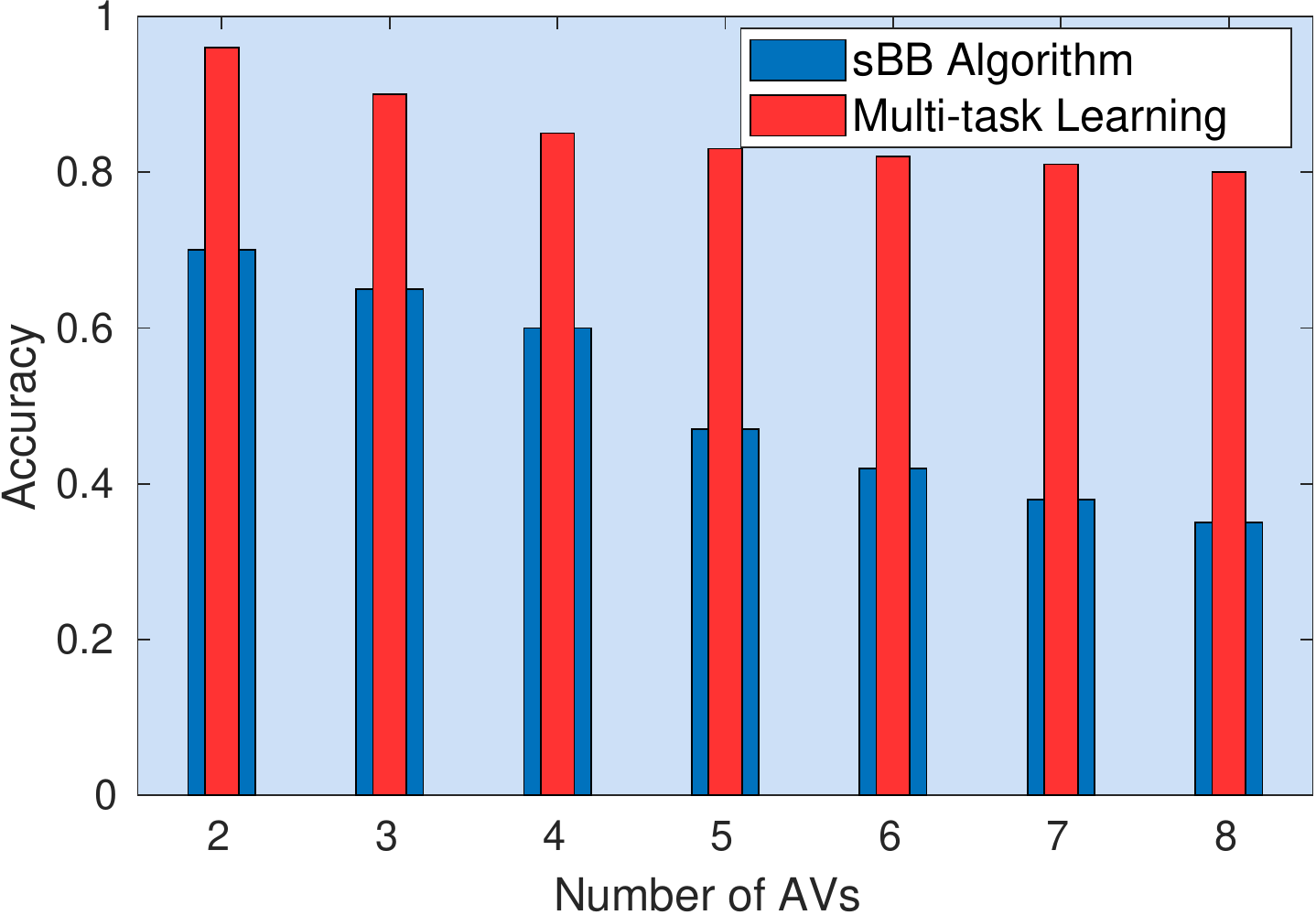}}
\caption{Inference accuracy of the classification and the MSE of the regression in the MTL model versus the percentage of training samples is shown in (a), where the number of AVs is $2$. In (b), the number of AVs versus the accuracy achieved by the sBB algorithm and the multi-task learning model is illustrated, respectively, where the percentage of training samples in multi-task learning model is 1.}
\label{fig:5}
\end{figure*}  

\subsection{Case Study: EI-Assisted Visual Object Detection}
In this section, we consider a visual object detection based autonomous driving system. Due to limited computing resources and a tight energy budget of self-driving vehicles, they usually need help from the BS to collaboratively process the delay-sensitive tasks. Since the offloading strategy decision-making with minimization of AVs' energy consumption and latency can be formulated as an MINLP problem, which is NP-Hard, we design an MTL-based feedforward neural network model by learning the end-to-end mapping between the problem input parameters and the output solutions. Specifically, the binary offloading decision is considered as a \textit{multi-class classification} problem with the cross-entropy loss, and the edge server computational resource allocation is considered as a \textit{regression} problem with the mean square error (MSE) loss. During the MTL training, the Adam optimizer is used to minimize the weighted-sum loss, where $\chi_c$ and $\chi_l$ denote their weights, respectively. 
 It should be noted that the size of the appropriately trained MTL model is less than $2$ KB, and thus the edge server has enough storage space to save the trained MTL model, which can be even cached into the edge server's memory in advance to perform the inference more efficiently. With the offline trained MTL model, the BS can directly infer the offloading strategy with high accuracy in near-real-time by performing feedforward calculation without iterations, as illustrated in step 1) - step 2) in Fig.~\ref{fig:4}. Therefore, the proposed MTL-based method ``moves" the complexity of online computation to offline training, which can be scaled up across the graphics processing unit (GPU) clusters. 

To achieve privacy-preserving and meet the hardware constraint of AVs, we investigate it by segmenting a trained CNN model into the lower-level SNN layers and the higher-level DNN layers. In particular, the SNN layers and DNN layers usually contain convolutional layers, rectified linear units layer, polling layer, and fully-connected layers. 
The SNN layers are deployed on the resource-constrained vehicles, usually with limited computation resources and storage space. All the AVs share the same DNN layers, which are deployed at the edge server and can further improve the inference accuracy by processing the feature maps generated by the SNN layers. This edge learning setup fulfills the demand for the timely processing of the video images while taking into account the practical implementation constraints, e.g., users' privacy. As illustrated in step 3) - step 4) in Fig.~\ref{fig:4}, the lower layers can directly provide sufficient features for the acceptable object detection performance without offloading when the captured images quality is sufficiently ``good". On the contrary, the captured images with ``bad" quality would offload the intermediate feature maps to the higher layers at the edge server to improve inference accuracy. The edge server then returns the processing results, which help the vehicles react fast and accurately via the ECUs.

\begin{figure}[t] 
            \centering
           \captionsetup{font={small }}
            \includegraphics[width=3.0in, height=2.4in]{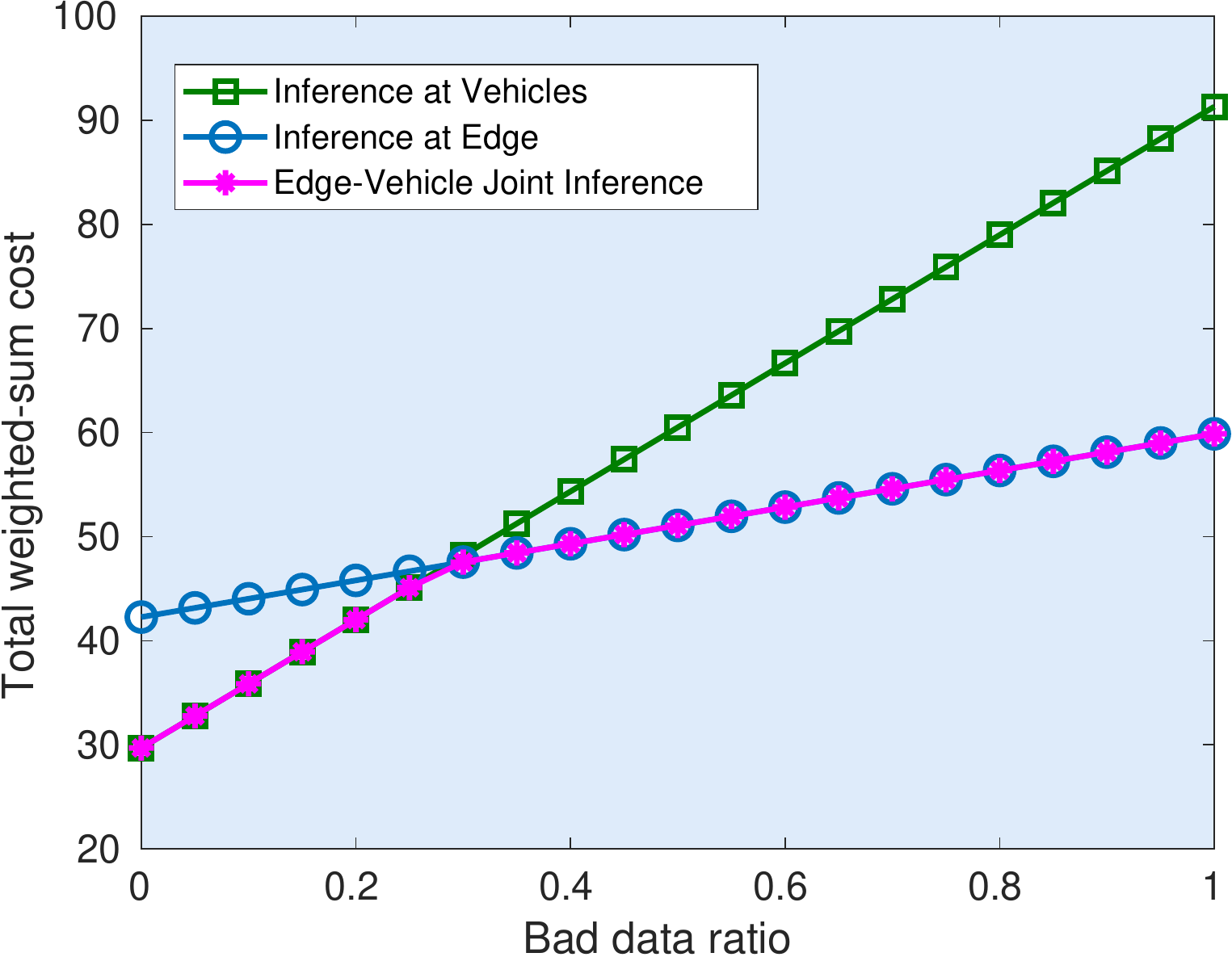} 
            \caption{Bad data ratio ($\eta$) versus total weighted-sum cost 0using three different kinds of inference methods, where the number of AVs is $2$, the transmission power of AVs is $10$ W, the CPU frequency of the AVs and the edge server is $1$ GHz and $10$ GHz, respectively.}
            \label{fig:6}
\end{figure}

\subsection{Illustrative Results} 
In this subsection, we first evaluate in Fig.~\ref{fig:5}(a) the impact of the number of training samples on the inference accuracy of the MTL model, where $\chi_c=\chi_r=1$. The inference accuracy is defined as the ratio of the number of correct predictions to the total number of predictions. We generate $4\times 10^4$ data samples by traversing the combinations of the binary offloading decision and computational resources allocation ratio with the exhaustive searching algorithm. In Fig. \ref{fig:5}(b), we compare the accuracy of the proposed MTL model with that of the conventional spatial branch and bound (sBB) algorithm, where $\chi_c=0$ and $\chi_r=1$ when the number of AVs is larger than $5$, and the percentage of training samples in MTL model is $1$. After, we investigate in Fig.~\ref{fig:6} the impact of the ``bad" data ratio (denoted as $\eta$) on the weighted-sum cost (defined as the weighted sum of the delay and energy consumption of all the AVs) using different inference strategies in the autonomous driving system. 

We observe from Fig.~\ref{fig:5}(a) that the inference accuracy of binary offloading decision (i.e., offload or not) is low for the small number of training samples and increases with the percentage of training samples, whilst the MSE of regression shows an opposite trend. The reason is that when the training samples are insufficient, the MTL model can only learn a few distinguishing features, which can hardly perform inference accurately. As the number of training samples increases, the MTL model can further extract more abstract features enabling the MTL model to learn more about the mapping between the input parameters and the output solutions of the formulated MINLP problem. From Fig~\ref{fig:5}(b), we observe that the MTL model always outperforms the sBB algorithm in the inference accuracy, and the gain is improved as the number of AVs increases. Compared to the sBB algorithm, almost $120\%$ improvement can be achieved by MTL model when the number of AVs is $8$. Notably, the MSE of the MTL model is even less than half of the sBB algorithm. Moreover, by using multi-task learning, the time consumption solving the MINLP is less than one-thousandth of the sBB scheme when the number of AVs varies from $2$ to $8$.

Fig.~\ref{fig:6} shows that when the ``bad" data ratio (denoted as $\eta$) becomes larger (e.g., more captured images are with low quality), this leads to an increase of the weighted-sum cost using three inference methods. Specifically, when $\eta$ is relatively small, e.g., $\eta<0.3$, inference at vehicles becomes more competitive due to the achieved high inference accuracy without introducing communication delay. As $\eta$ increases, e.g., $\eta \ge 0.3$, the inference given by the SNN layers becomes worse, which leads to a sudden rise in the weighted-sum cost due to the introduced penalty. At this time, the advantage of edge inference is an explicit benefit of exploiting more powerful DNN layers to improve inference accuracy via offloading the intermediate feature maps.

\section{Conclusion and Open Research Topics}
\label{conclusion}

In this article, we discussed the potential and challenges of edge intelligence in connected autonomous driving. We presented a two-tier edge learning-empowered architecture of wireless computing for AVs and introduced two functionality designs to make offloading strategy decision efficiently, ensure data privacy, and improve inference accuracy while meeting the delay constraint. The effectiveness of the proposed framework was demonstrated via a case study and experiments.

There are some interesting open research topics for EI-based autonomous driving architecture design. 
\begin{itemize}
\item \textit{Efficient edge learning with limited labeled training data}. Generally, most of the edge learning tasks in the autonomous driving system are based on supervised learning, which requires complete labeled training datasets. However, collecting sufficient labeled data with a correctness guarantee is still a challenge in practice~\cite{Ye}. To facilitate learning on edge with limited labeled data, we can intuitively take advantage of the historical data and combine it with real-time collected data. Also, we can take advantage of semi-supervised learning, transfer learning, or even autonomous learning~\cite{Ye}, to yield significant training and inference improvements via limited training data. Furthermore, the sensing deficiency caused by using only video cameras can be filled up by exploring the potential of sensor fusion, such as LiDAR sensors, radar sensors, and even by querying the sensor data from other vehicles. 
 
\item \textit{Distributed model training with restricted wireless bandwidth.} 
In the autonomous driving system with edge learning, the training data involved is generally privacy-sensitive with large size and quantity, e.g., high-definition video sequences. By introducing distributed learning (e.g., federated learning), only the local model parameters are uploaded to the edge server to preserve data privacy. Interestingly, via the federated learning and the neuron network segmentation, privacy-preserving can be achieved at the training side and inference side, respectively.
Since each AV usually collects unique training data separately and the data samples are generally non-independent and identically distributed among the vehicles, the edge server prefers to include more AVs' local FL models to generate a converged global model. In this context, the model parameters exchange between AVs and servers may result in high communication costs, and the runtime of each learning iteration is dominated by the slowest participants, which become a bottleneck for the autonomous driving system~\cite{ZY_TVT}. To mitigate this issue, more AVs can be allowed to participate in the FL by sharing their FL models with selected AVs via V2V communications, using the side-link channel in cellular-based V2V (C-V2V), DSRC, or even millimeter-wave transmission~\cite{{self-driving-survey},{XK_ITS}}.

\item \textit{Vehicle platoon-aided inference}.
In practice, the vehicles may sometimes drive out of the coverage of BS, which will cause considerable influence on the feasibility of edge inference. To address this challenge, collaborative caching and computing among vehicles become a competitive candidate solution~\cite{{ZY_TVT2},{XK_TVT}}. Specifically, the trained DNN model can be divided into multiple parts (called layers), each of which is pre-deployed at the vehicle belonging to the same vehicle platoon. Once a vehicle leaves the BS's coverage, this vehicle can still achieve reasonable inference by routing the intermediate parameters to other vehicles within the same vehicle platoon via V2V communications. In this case, the improvement of inference accuracy is at the expense of V2V communication delay, and the impact of the platoon dynamics needs to be further investigated.
\end{itemize}

\end{document}